\documentstyle[12pt]{article}
\begin{document}
\begin{flushright}
NSF-ITP-92/142\\
TPI-MINN-92/58\\
November 1992\\
\end{flushright}
\begin{center}
{\bf Nonperturbative QCD Effects in Weak Nonleptonic Decays}
\thanks{Talk  at DPF-92 meeting}
\end{center}
\begin{center}
{\bf B. Blok} \\
Institute for Theoretical Physics\\
University of California, Santa Barbara, CA 93106\\
and\\
{\bf M. Shifman} \\
Theoretical Physics Institute\\
University of Minnesota,
 Minneapolis, MN 55455\\
\end{center}
\begin{center}
{\bf\large Abstract}
\end{center}
{QCD-based analysis of nonfactorizable parts of weak nonleptonic
amplitudes is reported.
 Nonperturbative  effects due to soft gluon exchange
play a key role
leading to the emergence of a dynamical rule of discarding $1/N_c$
corrections. }
\section{Introduction}
\par Factorization is used in nonleptonic decays from early sixties. However,
as our knowledge of QCD and weak decay phenomenology deepens, the simple idea
that one must factorize two V-A currents composing
the effective weak hamiltonian evolves towards a rather sophisticated scheme
with different ingredients.
The purpose of this talk is to review  recent progress in calculating
deviations from naive factorization. We shall concentrate here on
exclusive decays.( Inclusive decays are discussed elsewhere.)
 We shall show that the rule of discarding $1/N_c$
\cite{1,2,3} has a dynamical origin, and is due to nonperturbative QCD
effects. The key role is played by soft chromomagnetic gluon exchange.
The resulting picture \cite{4,5,6,7,8}
is rather versatile$-$not all transitions are alike in this respect.
  QCD effects lead to
 deviations from the naive factorization, specific for each channels.
These deviations can be estimated in a model-independent way.
 In some channels the situation is close to the predictions of the  $1/N_c$
rule
, in others $-$ to naive factorization.  The degree of
cancellation of the naive $1/N_c$- suppressed amplitudes is different for each
channel, so we can call our approach a dynamical rule of discarding $1/N_c$.

---------------------------------------------------------------
\par $^1$ Talk  at DPF-92 meeting, FERMILAB, November 10-14, 1992.
\section{The method}
\par We shall describe our method using the decay
$B^0\rightarrow D^+\pi^-$ as an example.  The reader is referred to ref.
\cite{4} for details. We are interested in the transitions induced by  the
color-octet times color-octet part of the weak
 Lagrangian ${\cal L}\sim(\bar c\Gamma^\mu t^ab)(\bar d\Gamma^\mu t^a u)$.
In order to calculate the transition amplitude, we start with the correlator
\begin{equation}
{\cal A}^\beta=\int d^4x<D\vert T\{{\cal L}(x),A^\beta\}\vert\bar B>e^{iqx}
\label{eq:1}
\end{equation}
where the axial current $A^\beta=\bar u\gamma^\beta\gamma^5d$ annihilates
 the pion. Two key steps are made in order to calculate the latter correlator:
we continue ${\cal A}^\beta$ into the Euclidean region $-q^2\sim Q^2\sim
1\quad GeV^2$, and write in this region (borrowing some ideas from the
QCD sum rule method \cite{11}) a sum rule for the amplitude $M_{\rm n.f.}$
governed by ${\cal L}$:
\begin{equation}
{\cal A}^\beta (Q^2)=M_{\rm n.f.}{f_\pi q^\beta \over q^2}+...
\label{eq:2}
\end{equation}
where $+...$ denotes the contribution of higher resonances produced by the
 axial
current.
Second, we calculate ${\cal A}^\beta$ using the Operator Product Expansion.
We immediately obtain
\begin{equation}
{\cal A}^\beta (Q^2)=-2i{1\over 8\pi^2}{q^\alpha q^\beta\over q^2}
<D\vert \bar c \Gamma^\mu t^ag\tilde G^a_{\alpha\mu}b\vert B>+...
\label{eq:3}
\end{equation}
where $+...$ denotes higher order power corrections and we retained only
the kinematical structure proportional to $q^\beta$.
Comparing  the latter two equations
and neglecting the dots
 we immediately obtain that the ratio of
 the nonfactorizable
and  $1/N_c$ naive factorizable parts of the amplitude is
\begin{equation}
r=-{m^2_{\sigma H}\over 4\pi^2f^2_\pi}.
\label{eq:4}
\end{equation}
Note the key distinction from the standard
QCD sum rule method: the matrix elements
are taken between hadronic states, not between vacuum states.
Using the methods of HQET \cite{10} it is easy to get for $m^2_{\sigma H}=
{3\over 4}(m^2_{B^*}-m^2_B)$.
\par It is instructive to emphasize the assumptions and approximations made in
eq. (4). First, we neglected the corrections  due to  operators with
 higher dimensions, and contamination with higher resonances. Strictly speaking
it is necessary to check that the corresponding window exists. This has not
been done yet, although  arguments in favor of  smallness
of the above corrections in a large class of transitions were given
in ref. \cite{4}. Second, and this is also important,  we started from
the theoretical limit where $M_B-M_D\ll{1\over 2}(M_B+M_D)$. Only in this
limit the expansion in eq. (3) goes in dimensions, not twists.
Moreover, in a number of cases the hadronic matrix element in eq. (3)
reduces to the known quantity in this limit. Otherwise we would have to
introduce an unknown function of recoil. Logarithmic corrections due to
anomalous dimensions are also not included so far (although in the
transitions considered in ref. \cite{4} they seem  to be unimportant).
We refer to ref. \cite{4} for the detailed discussion of the method and
expected uncertainties. The expected accuracy for this particular channel
is of order one.
\par Keeping in mind all these uncertainties$-$a vast field for future work$-$
one can try to extend the method
 to other weak hadronic decays in a straightforward
way.  If the particle that splits away is not a pion, we do not
get a simple $1/Q^2$ term in the OPE, but
rather a
 more complicated function.
Moreover, the higher power corrections can become more important.
For example, for the $B\rightarrow DD$ decays we get
the function
$F(Q^2)=1/Q^2-{m^2\over Q^4}{\rm ln}({Q^2+m^2_c\over m^2_c})$
as a coefficient in front of the operator $G_{\mu\nu}$, instead of $1/Q^2$.
The relevant sum rule takes the form
\begin{equation}
{m^2_{\sigma H}\over 4\pi^2}F(Q^2)+... =f_D{M_{\rm n.f.}\over Q^2+m^2_D}+..
\label{eq:5}
\end{equation}
(where we once again neglected higher power corrections.)
It works well for the Euclidean momenta $Q^2\ge 1$
GeV$^2$.
\par The amplitudes of decays
   considered above were proportional to $a_1$ in the BSW
language \cite{1}.  The amplitudes of  decays proportional to $a_2$ contain
 an absolutely  unknown formfactorfactor,
the matrix element
$<B\vert \bar b\gamma^\nu g\tilde G_{\alpha\nu}u\vert\pi>$,
which cannot be detemined using HQET.
The sum rules for the decay $B^0\rightarrow D^0\pi^0$ and other decays of
 the type
"$B\rightarrow D$+light meson" in this group will be similar to the above,
(with the function $F(Q^2$) instead of $1/Q^2$)
 but will include  this  new
formfactor. For the decay $B\rightarrow J/\psi K$ we  have a new function
$\tilde F(Q^2)=2m^2_c\int^{\omega_c}_{4m^2_c}{ds\over (s+Q^2)\sqrt{s(s-4m^2)}}$
in the sum rule instead of F.
\section{ Decay widths}
Let us briefly discuss  numerical aspects of our results.
We shall concentrate on the values of $r$ and the amplitudes $a_1$ and
$a_2$ that can be directly compared with the experimental data.
\par  For decays  $B^0\rightarrow D^+\pi^-$
, $B^0\rightarrow D^+\rho^-$ we get $r\sim -1.5$ and $\sim -1$ respectively.
For the decays $B^0\rightarrow D^{*+}\pi^-$
, $B^0\rightarrow D^{*+}\rho^-$ we get $r'=r/3$ \cite{4}. Taking here and
below $c_1\sim 1.12,c_2\sim -0.26$, we obtain for these decays
$a_1$ $\sim$ 1.16, 1.12, 1.08, 1.06 respectively.
\par  Consider now other decays using the same method.
The discussion below is given for orientation only, keeping
in mind that the effects unaccounted for in our analysis (see section 2)
may be important for these decays.
If we neglect these effects, we obtain for them once again the
formulae similar to the one in eq. (3).
 Consider first the decays from the $B\rightarrow DD$ group.
 Their amplitudes are
also proportional to $a_1$ and can be obtained using the sum rule
sketched in  section 2. We get for $B\rightarrow DD$ decays
$r\sim -0.9{m^2_{\sigma H}\over 4\pi^2f^2_D}$, where $f_D$ is a leptonic decay
constant taken to be $\sim 170$ MeV. We immediately see that $r\sim -0.8$.
For $B\rightarrow D^*D^*$ decays using HQET we obtain $r\sim
-0.9{m^2_{\sigma H}\over 12\pi^2f^2_{D^*}}\sim -0.16$. For $B\rightarrow
D^*D$ decays we get $r\sim -0.5$. For the corresponding $a_1$ factors
in the amplitudes we find  $a_1\sim$ 1.1, 1.04, 1.07 respectively.
Note that our results for different channels lie between BSW \cite{1} and
naive factorization. The accuracy of these results is lower
than for the previous group of decays, since we expect here
the perturbative logs and higher corrections can play a bigger role.
\par Consider now the decays proportional to the factor $a_2$. Here we shall
be extremely speculative, since the
corresponding analysis is far from being completed.
We shall only try to indicate what we expect for these decays at the
moment, leaving  more solid statements for the future investigation.
The main difficulty here is that we do not know the key
formfactor$-<B\vert \bar bg\tilde G_{\alpha\mu}\gamma^\mu u\vert \pi>P^\alpha$
(and the corresponding formfactor with the $\rho$-meson). We can
 try to roughly estimate these formfactors from our
knowledge of the  D meson decays
using the symmetry between b and c.
 (Unfortunately, such estimates are very
uncertain, though.)  Let us completely ignore the recoil dependence in the
formfactor $<B\vert \bar bg\tilde G_{\alpha\mu}\gamma^\mu u\vert \pi>$
(unlike $B^0\rightarrow D^+\pi^-$ there is no justification for that)
and parametrize $<B\vert \bar bg\tilde G_{\alpha\mu}\gamma^\mu u\vert \pi>$
by a number
 $m^{'2}_{\sigma H}\sim x m^2_{\sigma H}$, where $x$ is an unknown constant.
 Then for
the decays of the type $B^0\rightarrow D^0\pi^0$ we obtain
 $r\sim -1.6x$,
 for  $B^0\rightarrow D^{*0}\pi^0$ we obtain
$r\sim -0.8x$ (the difference between the values of $r$
for decays to $D$ and $D^*$ is proportional to $f^2_D/f^2_{D^*}$
and we use $f_D=$ 170 MeV, $f_D^*$=220 MeV). The value of $x$ is not
known, but the experimental data on D seems to indicate that it is below 0.4.
If this is indeed the case for B decays,
then the value of
 $a_2$  will be
 suppressed in comparison with the exact $1/N_c$ rule for this group
of B decays, and can  even be equal to zero for sufficiently small $x$.
Such a suppression is favored by the recent experimental data \cite{9}.
 Future calculations
are needed to establish $x$, and at moment we cannot make any definite
theoretical statement about this group of decays.
\par Finally, we  note that the same calculation for $B\rightarrow
J/\psi K$ leads to small $r$ due to a big leptonic decay
constant $F_{J/\psi}\sim 300 $ MeV,
$a_2\sim 0.12$.
 However in this case there exist
new difficult problems, due to a large  recoil, an
enhanced role of higher power
corrections and higher twists and big continuum contribution (
presumably absent for other modes). Moreover, hard gluons can play
a significant role in this decay.
(The sum rule from section 2 has no stability "window" in this case).
Thus, we cannot exclude the possibility of the rule of discarding $1/N_c$
in this channel yet, neither can we confirm it.
\par  We   stress here that the pattern  of amplitudes proportional to
 $a_2$ presented above is  speculative and is nothing else than an educated
guess.
A lot of work, especially on the determination of
chromomagnetic nondiagonal formfactors remains to be done.
\section{Conclusion}
We tried here to draw a general picture for deviations from the naive
 factorization in B decays which stems from nonperturbative QCD.
A few aspects requiring further clarification are as follows. Higher
power corrections must be calculated, perturbative logs must be taken
into account and the nondiagonal magnetic formfactors must be determined.
After all this is done the expected accuracy of our results may be 20-30$\%$.
\par We also considered $K\rightarrow\pi\pi$ and $K-\bar K$ mixing parameter
(see ref. \cite{5}), as well as inclusive widths of B and D \cite{6,7,8}.
\par Summarizing, the  nonperturbative QCD effects play a key role
here in  weak hadronic decays and can be
responsible for the dynamical rule of discarding $1/N_c$.
We now have a general method that allows one to carry out these calculations
with reasonable accuracy for all decays of B,D and K.

\end{document}